\documentclass[twocolumn,showpacs,preprintnumbers,amsmath,amssymb]{revtex4}
\usepackage{graphicx}
\usepackage{dcolumn}
\usepackage{bm}

\begin{document}

\title{Asymmetric resonance in selective reflection: Explanation via Fano-like mechanism}

\author{Denis V. Novitsky}
 \email{dvnovitsky@tut.by}
\affiliation{%
B.I. Stepanov Institute of Physics, National Academy of Sciences of
Belarus, \\ Nezavisimosti~Avenue~68, 220072 Minsk, Belarus.
}%

\begin{abstract}
Fano mechanism is the universal explanation of asymmetric resonance
appearing in different systems. We report the evidence of Fano-like
resonance in selective reflection from a resonant two-level medium.
We draw an analogy with the asymmetric resonance previously obtained
for the coupled oscillators. We also take into account the effects
of dielectric background and local-field correction and connect our
results with optical bistability.
\end{abstract}


\maketitle

From the first observation of asymmetric resonance in the beginning
of the twentieth century, it was observed and studied in different
physical systems. Many works of the last fifty years were inspired
by the pioneering investigation of Ugo Fano \cite{Fano} who proposed
the universal mechanism of asymmetric resonances appearance. The
essence of this mechanism is the process of destructive and
constructive interference of two paths along which the system can
evolve. One of these paths concerns the usual symmetric
(Breit-Wigner) resonance, while the second one goes through the
broad continuum. In other words, there is the superposition of
discrete and continuum states. This superposition, or interference,
can be obtained in different classical and quantum systems leading
to the characteristic resonance with asymmetric profile known as
Fano resonance.

Fano resonances were widely studied previously in terms of atomic,
optical, and condensed-matter physics. The last decade was marked by
enormous interest to Fano-like behavior of photonic, plasmonic, and
nonostructured materials (see, for example, the recent reviews
\cite{Mirosh, Luk} and references therein). In this Letter we
analyze the asymmetric resonance in selective reflection of the
medium and make an attempt to explain it in terms of Fano
interference.

Selective reflection is a powerful method to investigate the
properties of quantum systems, such as atomic gases. The usual
scheme implies the study of reflection from an interface between a
gas and a solid window. The main difficulty is the necessity to take
into account the motion of atoms (Doppler effect) and the
atom-surface van der Waals interactions \cite{Nienhuis, Wang,
Bloch}. Other examples of asymmetric resonances were observed in
studies of magnetically induced polarization switching
\cite{Parriger} and double optical resonance \cite{Whitley}.

In this Letter we consider a simple model of asymmetric reflection
spectrum of atomic medium. This model allows to analyze the
mechanism of asymmetric response in pure form. The system under
consideration consists of a collection of two-level atoms embedded
in dielectric medium, so that atomic motion is absent. As an example
of such system one can take a glass or semiconductor sample doped
with quantum dots (artificial atoms). The outer medium can be air so
that the atom-wall interactions can be neglected. The starting point
of the model is the Bloch equations for the inversion $W$ and the
atomic polarization $R$ \cite{Bowd93, Cren08},
\begin{eqnarray}
\frac{\partial R}{\partial t}&=& -i \frac{\mu}{2 \hbar} \ell E W + i (\Delta-\ell \epsilon W) R - \gamma_2 R, \label{dPdt} \\
\frac{\partial W}{\partial t} &=& -i \frac{\mu}{\hbar} (\ell^* E^* R
- \ell E R^*) \nonumber \\ &-& 2 i (\ell^* - \ell) \epsilon |R|^2 -
\gamma_1 (W-W_{\textrm{eq}}), \label{dNdt}
\end{eqnarray}
where $E$ is the amplitude of macroscopic electric field, $\mu$ is
the transition dipole moment, $\Delta$ is the detuning of radiation
from resonance, $\gamma_1$ and $\gamma_2$ are the population and
polarization relaxation rates respectively, $W_{\textrm{eq}}$ is the
value of inversion at equilibrium. The parameter  $\epsilon=4 \pi N
\mu^2 / 3 \hbar$ is responsible for Lorentz shift due to local-field
correction, $N$ is the density of two-level atoms per unit volume,
$\hbar$ is the Planck constant. Here $\ell=(\varepsilon_d+2)/3$ is
the local-field enhancement factor due to polarizability of host
material with dielectric constant $\varepsilon_d=n_d^2$ (generally,
complex, however we restrict ourselves to nonabsorbing dielectrics).

In stationary regime the dependence of inversion on light intensity
is described by the cubic equation
\begin{eqnarray}
(W-W_{\textrm{eq}}) \left| i (\delta-\ell b W)-1 \right|^2+W I=0.
\label{eqinv}
\end{eqnarray}
Here $I=\mu^2 |\ell E|^2/\hbar^2 \gamma_1 \gamma_2$,
$\delta=\Delta/\gamma_2$ and $b=\epsilon/\gamma_2$ are the
normalized parameters.

Equation (\ref{eqinv}) has three solutions which take on real values
in different ranges of parameters, which results in appearance of
intrinsic optical bistability at certain conditions \cite{Hopf,
BenAryeh, Friedberg}). Therefore, one has to link these solutions to
obtain the full spectrum. In order to simplify, we consider the case
of absence of the local-field Lorentz shift at first. This
corresponds to quite low values of $\ell b$, below the threshold of
bistable response. Further we will analyze the influence of Lorentz
shift on the resonant reflection. So, neglecting the term $\ell b W$
in Eq. (\ref{eqinv}), we obtain the solutions of the Bloch equations
as follows,
\begin{eqnarray}
W=\frac{W_{\textrm{eq}} (\delta^2+1)}{\delta^2+1+I}, \label{invers}\\
R=\frac{\mu \ell W_{\textrm{eq}} (\delta-i)}{2 \hbar \gamma_2
(\delta^2+1+I)} E. \label{polar}
\end{eqnarray}

Now we can calculate the dielectric function of the complex medium
containing two-level atoms inside a dielectric as
\begin{eqnarray}
\varepsilon=1+4 \pi P/E, \label{epstot}
\end{eqnarray}
where the polarization is given by \cite{Cren08}
\begin{eqnarray}
P=\frac{\varepsilon_d-1}{4 \pi} E + \ell P_{\textrm{res}}.
\label{polartot}
\end{eqnarray}
Here $P_{\textrm{res}}=2 N \mu R$ is the nonlinear polarization due
to resonant atoms. Substituting the stationary solutions
(\ref{invers}) and (\ref{polar}), the final expression is
\begin{eqnarray}
\varepsilon=\varepsilon_d + \frac{3 \ell^2 b W_{\textrm{eq}}
(\delta-i)}{\delta^2+1+I}. \label{epstot1}
\end{eqnarray}

Let us consider a monochromatic electromagnetic wave incident
normally to the interface between the outer medium with the
refractive index $n_0$ and two-level medium with the dielectric
permittivity (\ref{epstot1}). Assuming the thickness of the latter
medium much less than the light wavelength, we can avoid the
necessity of taking into account the propagation effects. Then
reflection of light by the interface can be described by the
amplitude coefficient as follows,
\begin{eqnarray}
r(\delta)=\frac{\sqrt{\varepsilon}-n_0}{\sqrt{\varepsilon}+n_0}.
\label{reflcoef}
\end{eqnarray}
Using Eq. (\ref{epstot1}), one can easily obtain the explicit
expression for the relation $r(\delta)$. However, we will not write
it out, since it is awkward and not illustrative. Further we just
demonstrate it in graphical form.

\begin{figure}[t!]
\includegraphics[scale=0.9, clip=]{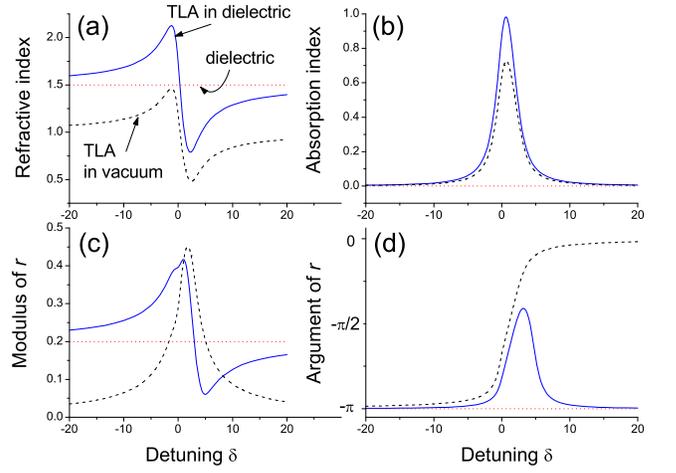}
\caption{\label{fig1} Spectral curves for (a) refraction index
$n=\textrm{Re} \sqrt{\varepsilon}$, (b) absorption index
$\kappa=\textrm{Im} \sqrt{\varepsilon}$, (c) absolute value of the
reflection coefficient, (d) its phase. The notation of figure (a) is
valid for (b), (c) and (d) as well: solid line refers to the
two-level atoms (TLA) in dielectric, dashed -- TLA in vacuum, dotted
-- dielectric itself. The parameters of calculation: $n_d=1.5$,
$n_0=1$, $b=1$, $I=1$, $W_{\textrm{eq}}=-1$ (these values are valid
for other figures if the other is not stated).}
\end{figure}

Figure \ref{fig1} shows spectral dependencies of refraction and
absorption indices, as well as of modulus and argument of the
reflection coefficient (\ref{reflcoef}). It is seen that the modulus
of $r$ follows the symmetrical curve in the case of two-level atoms
in vacuum, while its phase demonstrates the phase jump by $\pi$ near
the resonance. The reflection from the dielectric itself is
described by continuum-like curves. The combination of two-level
atoms and dielectric results in fundamentally another behavior:
Asymmetric line-shape for modulus, and the gain and the subsequent
drop for the phase of the reflection coefficient. This behavior is
similar to that of the excited oscillator coupled to another one
\cite{Joe}. We treat this behavior as the evidence of Fano resonance
in reflection from two-level atoms embedded in a dielectric medium.

\begin{figure}[t!]
\includegraphics[scale=0.9, clip=]{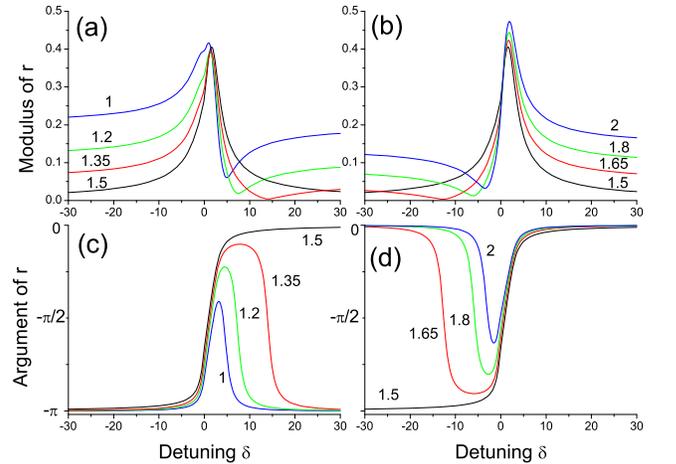}
\caption{\label{fig2} Spectral curves for (a, b) the absolute value
of the reflection coefficient, (c, d) its phase at different values
of refractive index of the outer medium $n_0$ (numbers near the
curves).}
\end{figure}

The role of asymmetry parameter plays the difference $\Delta
n=n_d-n_0$ between the refractive indices of the host dielectric and
the outer medium. This is seen from Fig. \ref{fig2}: At $\Delta n=0$
we have a symmetric resonance, while at nonzero values of this
difference the asymmetric curves are observed with minimal values of
reflection on the left or on the right of $\delta=0$ depending on
the sign of $\Delta n$. However, the situation is more complex:
Comparison of Figs. \ref{fig2}(a) and (b) shows that curves at the
same $|\Delta n|$ and different signs are not exactly the same. It
is also seen that the value of reflection at minimum is not zero and
gets larger as the difference $\Delta n$ is increasing. Continuing
our analogy, one can say that, in the case of coupled oscillators,
the curve cannot reach zero as well if only the frictional parameter
of the second oscillator is not zero \cite{Joe}.

\begin{figure}[t!]
\includegraphics[scale=0.9, clip=]{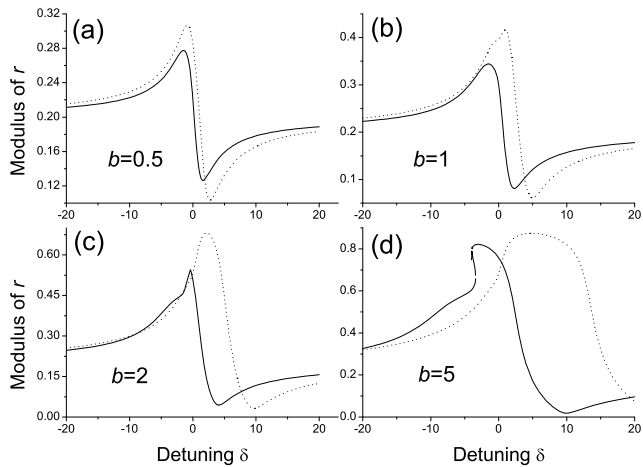}
\caption{\label{fig3} Spectral curves for the absolute value of the
reflection coefficient at different values of parameter $b$: (a)
$b=0.5$, (b) $b=1$, (c) $b=2$, (d) $b=5$. Solid lines correspond to
the calculations taking into account of local field effect; dotted
lines were calculated using Eq. (\ref{epstot1}).}
\end{figure}

Now let us consider the effect of local field on the asymmetric
resonance. The description of medium is then given by the same
equations (\ref{epstot}) and (\ref{polartot}), where the microscopic
polarization is
\begin{eqnarray}
R=\frac{1}{2} \frac{i \ell W \Omega}{i (\delta-\ell b W)-1},
\label{polar}
\end{eqnarray}
and inversion is governed by the cubic equation (\ref{eqinv}).
Comparison between this description and those by Eq. (\ref{epstot1})
is presented in Fig. \ref{fig3}. It is seen that influence of local
field is not large at low values of coefficient $b$ which is often
referred to as the near-dipole-dipole interactions parameter. But as
$b$ increases the difference between solid and dotted lines in Fig.
\ref{fig3} grows: the latter curve shifts towards larger detunings
that is compensated by Lorenz shift in the former case. At the same
time the amplitude of the resonance tends to grow and its minimum
reaches closer to zero. Another important effect which appears due
to the local field is the bistable response. One can see it in Fig.
\ref{fig3}(d) near $\delta=-4$ (solid line) while the dotted curve
does not have such feature. Finally, it is worth noting that local
field effects (such as optical bistability) can be enhanced due to
the host dielectric with large refraction index \cite{Nov11}.

\begin{figure}[t!]
\includegraphics[scale=0.85, clip=]{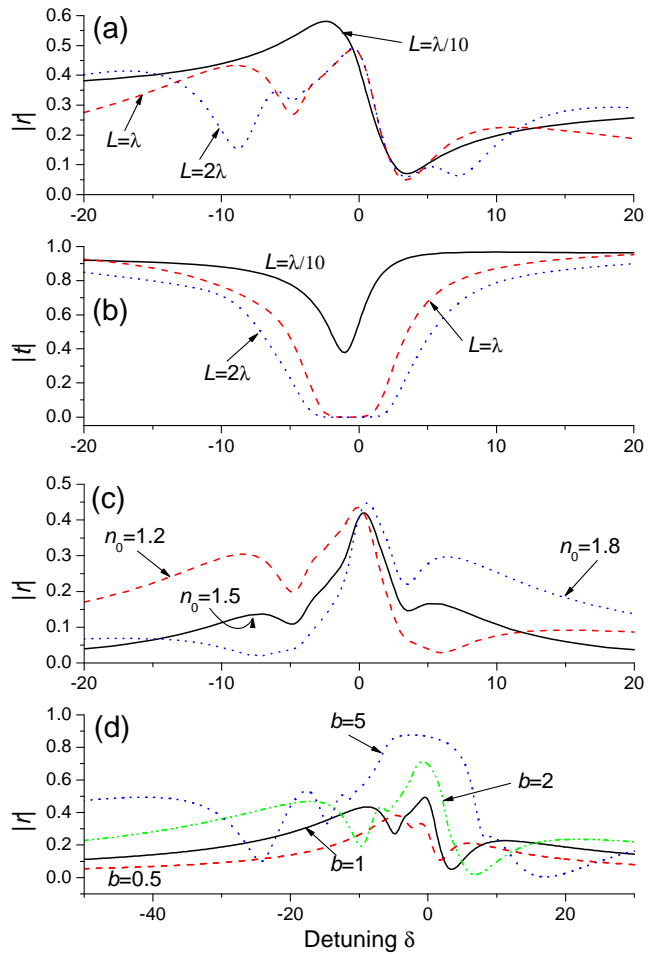}
\caption{\label{fig4} Spectral curves for the absolute value of (a,
c, d) reflection coefficient, (b) transmission coefficient. The
calculations were performed for the layers of different (a, b)
thickness (in units of the light wavelength $\lambda=0.5$ $\mu$m);
(c) outside refractive index $n_0$ (at $b=1$, $L=\lambda$), (d)
parameter $b$ (at $n_0=1$, $L=\lambda$). Note that local field
correction is taken into account.}
\end{figure}

The results considered above were obtained for the local interface
between the outer medium and the two-level medium in the dielectric
host. Further we try to verify them for more general and realistic
case of the resonant medium of finite thickness. Our calculations
were performed within the framework of the iteration matrix method
discussed in detail previously \cite{Nov08}. This approach is
sufficient to obtain stationary characteristics, though, in general,
one needs to use the coupled Maxwell-Bloch equations to study light
propagation \cite{Bowd93}. Figure \ref{fig4} shows that, for the
layers thin in comparison with the radiation wavelength (for
example, $L=\lambda/10$), we have qualitatively the same asymmetric
resonance as in local case (Fig. \ref{fig1}). However, if the the
thickness of the layer is about or larger than $\lambda$, some local
minima appear in the spectral curves of the reflection coefficient.
These additional minima are due to propagation effects, that is
inhomogeneous distribution of the medium properties (refractive
index) along layer's thickness. At the same time, the dip in
transmission is getting wider for thicker layers due to large
absorption.

As demonstrates Fig. \ref{fig4}(c), the difference $\Delta n$ can
still be considered as the Fano asymmetry parameter though even for
$\Delta n=0$ the curve of resonance (solid line) is distorted
because of propagation effects. These distortions also grow up as
the local-field parameter $b$ increases [Fig. \ref{fig4}(d)].

In conclusion, we considered a simple model of selective reflection
from a resonant medium which consists of two-level atoms embedded in
a dielectric host. The asymmetric curve of the reflection
coefficient can be explained in terms of Fano resonance with the
difference between refractive indices of outer and host media as the
asymmetry parameter. This model can be considered as one more
evidence of universality of Fano mechanism alongside with classical
coupled resonators \cite{Joe} and recently reported optically driven
atomic force microscope cantilever \cite{Kadri}.


\begin{thebibliography}{15}
\bibitem{Fano} U. Fano, Phys. Rev. {\bf124}, 1866 (1961).
\bibitem{Mirosh} A. E. Miroshnichenko, S. Flach, and Yu. S. Kivshar, \rmp {\bf82}, 2257 (2010).
\bibitem{Luk} B. Luk'yanchuk, N. I. Zheludev, S. A. Maier, N. J. Halas, P. Nordlander, H. Giessen, and C. T. Chong, Nat. Materials {\bf9}, 707 (2010).
\bibitem{Nienhuis} G. Nienhuis, F. Schuller, and M. Ducloy, \pra {\bf38}, 5197 (1988).
\bibitem{Wang} P. Wang, A. Gallagher, and J. Cooper, \pra {\bf56}, 1598 (1997).
\bibitem{Bloch} D. Bloch and M. Ducloy, Adv. At. Mol. Opt. Phys. {\bf50}, 91 (2005).
\bibitem{Parriger} C. Parriger, P. Hannaford, and W. J. Sandle, \pra {\bf34}, 2058 (1986).
\bibitem{Whitley} R. M. Whitley and C. R. Stroud, Jr., \pra {\bf14}, 1498 (1976).
\bibitem{Bowd93} C.M. Bowden and J.P. Dowling, \pra {\bf47}, 1247 (1993).
\bibitem{Cren08} M. E. Crenshaw, \pra {\bf78}, 053827 (2008).
\bibitem{Hopf} F. A. Hopf, C. M. Bowden, and W. H. Louisell, \pra {\bf29}, 2591 (1984).
\bibitem{BenAryeh} Y. Ben-Aryeh, C. M. Bowden, and J. C. Englund, \pra {\bf34}, 3917 (1986).
\bibitem{Friedberg} R. Friedberg, S. R. Hartmann, and J. T. Manassah, \pra {\bf40}, 2446 (1989).
\bibitem{Joe} Y. S. Joe, A. M. Satanin, and C. S. Kim, Phys. Scr. {\bf74}, 259 (2006).
\bibitem{Nov11} D. V. Novitsky, \josab {\bf28}, 18 (2011).
\bibitem{Nov08} D. V. Novitsky, \josab {\bf25}, 1362 (2008).
\bibitem{Kadri} S. Kadri, H. Fujiwara, and K. Sasaki, Opt. Express {\bf19}, 2317 (2011).
\end{thebibliography}
\end{document}